\documentclass[reqno,11pt]{amsart}
\usepackage[margin=1in]{geometry}
\usepackage{amsmath,amssymb,amsthm,amsfonts}
\usepackage{graphicx}
\usepackage{booktabs}
\usepackage{multirow}
\usepackage{enumitem}
\usepackage{mathtools}
\usepackage{etoolbox}
\usepackage{hyperref}
\hypersetup{hidelinks}
\usepackage{array}
\usepackage{float}

\newtheorem{remark}{Remark}[section]
\numberwithin{equation}{section}

\makeatletter
\patchcmd{\@settitle}{\uppercasenonmath\@title}{}{}{}
\def\@settitle{%
  \begin{center}%
    \baselineskip14\p@\relax \large\bfseries \@title
  \end{center}%
}
\makeatother
\makeatletter
\def\@setauthors{%
  \begingroup \trivlist
  \centering\footnotesize \@topsep30\p@\relax
  \advance\@topsep by -\baselineskip
  \item\relax
  \andify\authors
  \def\baselinestretch{1}%
  \authors
  \endtrivlist \endgroup }
\makeatother

\newcommand{\dx}{\Delta x}

\newcommand{\pstar}{p_\ast}
\newcommand{\ustar}{u_\ast}

\newcommand{\wdot}{\mathring{w}}
\newcommand{\zdot}{\mathring{z}}
\newcommand{\sdot}{\mathring{s}}

\title[Sub-grid wave correction from DRVs]%
  {Intermittent  Sub-grid Wave Correction from Differentiated Riemann Variables}
\author[Steve Shkoller]{{\large Steve Shkoller}}
\date{}

\begin{document}

\begin{abstract}
We introduce a low-cost every-$K$-step correction for one-dimensional Euler computations. The correction uses
differentiated Riemann variables (DRVs)---characteristic derivatives that isolate the left acoustic wave, the contact,
and the right acoustic wave---to locate the current wave packet, sample the surrounding constant states, perform a short
Newton update for the intermediate pressure and contact speed, and conservatively remap a sharpened profile back onto
the grid. The ingredients are elementary---filtered centered differences, local state sampling, a single Newton step,
and conservative cell averaging---yet the effect on accuracy is disproportionate. On a long-time severe-expansion
benchmark ($N=900$, $t=0.4$), intermittent correction drives the intermediate-state errors from $O(10^{-2})$ to
$O(10^{-13})$, i.e.\ to machine precision. On a long-time LeBlanc benchmark ($N=800$, $t=1$), the method crosses a
qualitative threshold: one-shot final-time reconstruction fails entirely (shock location error $2.7\times 10^{-1}$),
whereas correction every three steps recovers an almost exact sharp solution with contact and shock positions accurate to machine precision. The same detector-and-Newton mechanism handles two-shock and two-rarefaction packets without case-specific logic,
with plateau improvements of four to sixteen orders of magnitude. In an unoptimized Python prototype the wall-clock
overhead is below a factor of two even on the most aggressively corrected benchmark. To our knowledge, no comparably
lightweight fixed-grid add-on has been shown to recover this level of coarse-grid accuracy on the long-time LeBlanc and
related near-vacuum problems.
\end{abstract}

\maketitle
\setcounter{tocdepth}{1}
\tableofcontents

\newpage

\section{Introduction}

The one-dimensional compressible Euler equations with Riemann initial data generate three fundamental wave families: a
left acoustic wave, a contact discontinuity, and a right acoustic wave. Depending on the data, the two acoustic waves
may be shocks or rarefactions, and between them lie constant intermediate states. Standard fixed-grid shock-capturing
methods often recover the qualitative wave ordering, but on strong problems they broaden the transitions, corrupt the
intermediate states, and gradually lose the sub-cell geometry needed for a sharp description of the solution.

The central idea of this paper is that this lost geometry need not be recovered only once, after the computation is
finished. It can be detected repeatedly during the run and fed back into the evolution itself. In our companion
paper~\cite{ShkollerDRV2026}, the same underlying characteristic information was used only at final time, as a
postprocessing diagnostic. Here it is used every $K$ steps to correct the evolving numerical solution.

The key diagnostic objects are the differentiated Riemann variables (DRVs), characteristic derivatives that separate the
left acoustic wave, the contact, and the right acoustic wave into different localized spikes. At a correction time,
filtered DRV surrogates identify the current wave packet, neighboring constant states are sampled from the resolved
solution, a short Newton update supplies the intermediate pressure and contact speed, and the resulting sharp profile is
conservatively remapped onto the grid. The method therefore does not replace the baseline solver; it provides occasional
geometric resets that keep the evolving cell averages close to the sharp solution the solver is trying to approximate.

The numerical effect is strikingly out of proportion to the simplicity of the correction. The headline results are as
follows.
\begin{itemize}[nosep]
\item \textsf{Long-time severe expansion} ($N=900$, $t=0.4$,
  $K=50$): intermediate-state velocity and pressure errors drop from $5.7\times 10^{-2}$ and $2.4\times 10^{-4}$ to
  $2.4\times 10^{-13}$ and $1.8\times 10^{-15}$---machine precision---with essentially no wall-clock penalty on this case.
\item \textsf{Long-time LeBlanc} ($N=800$, $t=1$, $K=3$):
  one-shot final-time reconstruction fails (shock misplaced by $2.7\times 10^{-1}$); intermittent correction recovers
  exact contact and shock positions (both errors at machine precision) and reduces the sharp-profile $L^1$ velocity error by
  four orders of magnitude.
\item \textsf{Two-shock collision and two-rarefaction expansion}:
  the same code path, with no case-specific logic, delivers plateau improvements of five to sixteen orders of magnitude
  on Toro's Test~4 and Test~2 initial data.
\end{itemize}
The cost is modest: in the present unoptimized Python prototype, the wall-clock overhead ranges from $0.93\times$
(faster, because the correction reduces total Euler steps) to $1.84\times$ on the most aggressively corrected benchmark.

Two additional tests extend the scope. A noninteracting Double-Sod calculation (Appendix~\ref{sec:doublesod-appendix}) shows that the method is not
restricted to a single global similarity center, and the illustrative near-vacuum problem in Appendix~\ref{sec:hypdeg-appendix} demonstrates the
sub-cell detection mechanism on a Mach $1.5\times 10^5$ benchmark where the contact and shock occupy less than one-fifth
of a single cell.

There are certainly one-dimensional methods that can produce very sharp discontinuities, including front tracking, shock
fitting, and fitted-front finite-element schemes \cite{LeVequeShyue1995,WitteveenKorenBakker2007,DAquilaHelenbrookMazaheri2020}.
Other stabilizing strategies, such as tuned artificial-viscosity closures \cite{ReisnerSerencsaShkoller2013}, are also relevant
historical context, but they are architecturally different from the present correction. The striking point here is different: an otherwise standard fixed-grid WENO-5/HLLC code,
augmented only by filtered centered differences, local state sampling, one or a few Newton updates, and conservative
remapping, can recover wave structures that are nearly exact from grids that would ordinarily seem far too coarse. On
the benchmarks reported below, that recovery is not cosmetic; on the long-time LeBlanc problem it changes a failed final
reconstruction into an almost exact one. To our knowledge, no comparably lightweight fixed-grid add-on has been shown to
recover this level of coarse-grid accuracy on the long-time LeBlanc and related near-vacuum benchmarks.

\subsection{Minimal notation and the DRVs}

For convenience, we record the basic definitions used throughout. We write the ideal-gas Euler equations in conservative
form
\[
\partial_t U + \partial_x F(U)=0, \qquad U=(\rho,\rho u,E)^T, \qquad F(U)=(\rho u,\rho u^2+p,u(E+p))^T,
\]
with
\[
p=(\gamma-1)\Bigl(E-\tfrac12\rho u^2\Bigr), \qquad c=\sqrt{\gamma p/\rho}, \qquad s=\log p-\gamma\log\rho
\]
(up to an irrelevant additive constant in $s$). As in~\cite{ShkollerDRV2026}, let
\[
\alpha=\tfrac{\gamma-1}{2}, \qquad \sigma=\tfrac{c}{\alpha}, \qquad w=u+\sigma, \qquad z=u-\sigma.
\]
The differentiated Riemann variables (DRVs) are then defined by
\begin{equation}
\label{eq:drv-defs-followup}
\wdot = \partial_x w
      - \tfrac{\sigma}{2\gamma}\partial_x s
      = \partial_x u
      + \tfrac{\alpha\sigma}{\gamma p}\partial_x p,
\qquad
\zdot = \partial_x z
      + \tfrac{\sigma}{2\gamma}\partial_x s
      = \partial_x u
      - \tfrac{\alpha\sigma}{\gamma p}\partial_x p,
\qquad
\sdot = \partial_x s.
\end{equation}
The formulas in \eqref{eq:drv-defs-followup} are evaluated in the code by centered-difference algebraic surrogates
computed from the current primitive fields and then adaptively Gaussian filtered before geometry detection. Numerically, $\wdot$ and
$\zdot$ carry the two acoustic families, while $\sdot$ carries the contact family. For a single-interface Riemann
packet, the two intermediate constant states separated by the contact are called the left and right \emph{star states};
together they form the \emph{star region}. The nearly constant values sampled there are the \emph{plateau values} used
by the algorithm. By a \emph{local Riemann packet} we mean the three-wave structure generated by a single interface
before it interacts with any neighboring packet. We also use the shorthand $1$-R/$2$-C/$3$-S, $1$-S/$2$-C/$3$-S, etc., to record the wave types of the
left acoustic wave, the contact, and the right acoustic wave.

\section{Intermittent DRV correction}

The baseline solver is the same as in our companion paper~\cite{ShkollerDRV2026}: componentwise WENO-5
reconstruction~\cite{JiangShu1996}, HLLC numerical flux~\cite{ToroSpruceSpeares1994}, and SSP-RK3 time
stepping~\cite{ShuOsher1989}. The new ingredient is an intermittent correction applied every $K$ steps.

Before stating the algorithm, we record the classical wave functions it uses. If $k\in\{L,R\}$ and $c_k=\sqrt{\gamma
p_k/\rho_k}$, then
\[
f_k(p)=
\begin{cases}
\dfrac{2c_k}{\gamma-1}
  \left[\left(\dfrac{p}{p_k}\right)^{%
  \tfrac{\gamma-1}{2\gamma}}-1\right],
  & p\le p_k,\\[2ex]
(p-p_k)\sqrt{\dfrac{2}{(\gamma+1)\rho_k
  \left(p+\tfrac{\gamma-1}{\gamma+1}p_k\right)}},
  & p>p_k,
\end{cases}
\]
with derivative
\[
f_k'(p)=
\begin{cases}
\dfrac{c_k}{\gamma p_k}
  \left(\dfrac{p}{p_k}\right)^{%
  -\tfrac{\gamma+1}{2\gamma}},
  & p\le p_k,\\[2ex]
\sqrt{\dfrac{2}{(\gamma+1)\rho_k
  \left(p+\tfrac{\gamma-1}{\gamma+1}p_k\right)}}
  \left(1-\dfrac{p-p_k}{2\left(p
  +\tfrac{\gamma-1}{\gamma+1}p_k\right)}\right),
  & p>p_k.
\end{cases}
\]
The pressure-wave-function equation is
\begin{equation}\label{eq:riemann-F}
F(\pstar)
  = f_L(\pstar) + f_R(\pstar) + u_R - u_L = 0,
\end{equation}
with $F'(p)=f_L'(p)+f_R'(p)$, and the recovered contact speed is
\[
\ustar=\tfrac12\bigl(u_L+u_R +f_R(\pstar)-f_L(\pstar)\bigr).
\]

At a correction time~$t^n$, with current conservative cell averages $(\bar\rho_j^n,\overline{\rho u}_j^n,\bar E_j^n)$ and
primitive values $(\rho_j^n,u_j^n,p_j^n)$ computed cellwise from them, the algorithm proceeds as follows.

\begin{enumerate}[leftmargin=*, nosep,
  label={\upshape\bfseries Step \arabic*.}]
\item Convert the current primitive variables into algebraic DRV
  surrogates $\wdot$, $\sdot$, $\zdot$ and a filtered positive part of~$u_x$.
\item Detect the rarefaction head and tail, the contact, and the
  shock from the DRV spikes and rarefaction support, thereby obtaining an approximate $1$-$2$-$3$ geometry
  \[
  X_{\mathrm{rh}}^n,\qquad X_{\mathrm{rt}}^n,\qquad X_c^n,\qquad X_s^n.
  \]
\item Sample left, left-star, right-star, and right plateau
  states from the current numerical profile.
\item Form the one-step approximate closure for $\pstar$ and
  $\ustar$ from the sampled seed
  \[
  \pstar^{(0)}=\tfrac12\bigl(p^{\mathrm{sample}}_{\ast,L} +p^{\mathrm{sample}}_{\ast,R}\bigr)
  \]
  followed by one Newton update of~\eqref{eq:riemann-F},
  \begin{equation}\label{eq:newton-step}
  \pstar^{(1)}=\pstar^{(0)}
    -\tfrac{F(\pstar^{(0)})}{F'(\pstar^{(0)})},
  \qquad
  \ustar^{(1)}=\tfrac12\bigl(u_L+u_R
    +f_R(\pstar^{(1)})-f_L(\pstar^{(1)})\bigr).
  \end{equation}
  Step~4, given explicitly by~\eqref{eq:newton-step}, is the local Riemann-informed correction.
\item Reconstruct the corresponding sharp self-similar profile at
  the \emph{current time} $t^n$ and replace the current cell averages by the cell averages of that sharp profile. Step~5
  is the feedback step.
\end{enumerate}

The dependence on the current time is essential. The correction applied at time~$t^n$ uses the wave speeds and plateau
states inferred from the current resolved solution and reconstructs the self-similar state at~$t^n$, not at the terminal
time. Once $(\pstar,\ustar)$ are known, the sharp local profile is piecewise constant on the plateau regions and uses
the usual self-similar simple-wave formula inside each detected rarefaction fan. More Newton steps can be taken whenever
a harder local closure is required.

\begin{remark}
In the present implementation each correction acts on one local Riemann packet with a known interface position. The main
text treats the standard $1$-R/$2$-C/$3$-S case together with $1$-S/$2$-C/$3$-S and $1$-R/$2$-C/$3$-R packets;
Appendix~\ref{sec:doublesod-appendix} records a noninteracting two-interface example whose right packet realizes the remaining local
$1$-S/$2$-C/$3$-R configuration. Truly post-interaction data and automatically emerging local packets remain outside the
present scope.
\end{remark}

\subsection{Intermittent Newton refinement}
\label{sec:distributed-newton}

When the correction is applied intermittently, the seed at correction $n+1$ is the $\pstar$ computed at correction~$n$,
stored in the reconstructed cell averages and recovered by plateau sampling at the next correction. The successive
one-step closures therefore behave like a \emph{distributed Newton refinement} across time steps. When the sampled
plateau states vary only mildly between successive corrections, only a few such updates are needed to drive $\pstar$
from a crude initial guess to near machine precision. For example, on the severe-expansion benchmark
(Section~\ref{sec:severe}), the star-state pressure reaches 13 significant digits within five corrections---five single
Newton updates, not fifty---and no exact local Riemann solve is called in the main algorithm.

\subsection{Distinction from the final-time postprocessor}

The intermittent algorithm is distinct from the final-time-only postprocessor of~\cite{ShkollerDRV2026}. The
final-time-only method changes only the output representation. By contrast, the intermittent algorithm replaces the
evolving cell averages and therefore feeds sub-cell information back into later time steps. This is the sense in which
it behaves like an evolution-time sub-grid correction: the sharp state injected at step~$n$ becomes the initial data for
steps $n+1,\ldots,n+K$.

\section{A deterministic, Riemann-informed sub-grid viewpoint}
\label{sec:les-analogy}

With the algorithm in hand, we can describe its conceptual position. The correction infers unresolved wave geometry from
the resolved field and feeds it back into the evolution---a structural pattern shared with large-eddy simulation (LES)
of turbulence, though the mechanism here is deterministic rather than statistical.

\subsection{Filtering and the correction problem}

In large-eddy simulation of turbulence, one begins with a filtered velocity field and then asks how the unresolved
eddies modify the resolved dynamics; see, for example,~\cite{Pope2000}. Here the starting point is different but the
structural question is similar. A shock-capturing Euler solver produces cell averages $(\bar\rho_j,\overline{\rho
u}_j,\bar E_j)$ in which discontinuities have been smeared over several cells by numerical diffusion. The intermittent
DRV algorithm asks whether the hidden sub-cell wave geometry can be inferred from those resolved data and then
reinserted into the subsequent evolution.

For the one-dimensional Riemann packets considered here, the answer is often yes. The filtered DRV surrogates identify
the local wave geometry, plateau sampling provides approximate outer and star states, and a short Newton update of the
pressure-wave-function equation supplies a local closure for the star pressure and contact speed. The resulting
sharpened profile is then averaged back onto the grid at the current time.

\subsection{A metaphorical comparison with LES}

The comparison with LES is heuristic rather than literal. In LES one filters the Navier--Stokes equations and introduces
a model for the effect of the unresolved eddy field on the resolved one. In the present Euler setting, one begins from a
shock-capturing computation and applies a local wave-packet correction every $K$ steps. The common feature is simply
this: unresolved structure is inferred from the resolved field and then fed back into later time steps.

Table~\ref{tab:les-analogy} summarizes this metaphorical comparison. The crucial difference is that the present
correction is not statistical. Its residual error comes from wave detection, plateau sampling, finite correction
frequency, and the use of a one-step Newton closure, not from an eddy-viscosity-type ansatz.

\begin{table}[ht]
\centering
\renewcommand{\arraystretch}{1.3}
\caption{{\footnotesize A metaphorical comparison between LES of turbulence
  and intermittent DRV correction.}}
\label{tab:les-analogy}
\begin{tabular}{p{3.6cm}p{5.0cm}p{5.3cm}}
\toprule
& \textsf{LES of turbulence}
& \textsf{Intermittent DRV correction} \\
\midrule
Governing equations
  & Navier--Stokes & Euler \\
Resolved field & Grid-scale velocity $\bar{u}_i$
  & Cell averages $(\bar{\rho},\overline{\rho u},\bar{E})$ \\
Unresolved structure & Turbulent eddies $<\dx$
  & Wave transitions $<\dx$ \\
Sensor & Filtered stress / gradients
  & DRV surrogates $\wdot$, $\sdot$, $\zdot$ \\
What is inferred & Net eddy effect on large scales
  & Wave locations, types, and local plateau data \\
Correction step & Statistical / phenomenological closure
  & One-step Newton update for local Riemann closure \\
Dominant source of residual error & Model mismatch
  & Detection, sampling, and approximate local closure \\
Feedback to resolved field & Modified stress or flux & Every-$K$-step replacement by a sharpened local profile \\
\bottomrule
\end{tabular}
\end{table}

\subsection{The correction is deterministic}

The fundamental difference from turbulence is the nature of the unresolved structure. In turbulence, the small scales
are chaotic and cannot be recovered deterministically from the resolved field alone. In a local Euler Riemann packet, by
contrast, once the outer states and wave pattern are fixed, the ideal-gas Riemann problem determines the intermediate
states. In the present implementation we do not call an exact local Riemann solver in the main algorithm; instead we use
one Newton update of the classical pressure-wave-function~\eqref{eq:riemann-F}, seeded by the sampled star-pressure
estimate. The point is that the local closure is Riemann-informed and deterministic, even though the algorithmic
realization remains approximate.

This distinction matters. The present paper does not claim a general LES closure for arbitrary Euler data. It shows,
rather, that for the one-dimensional Riemann packets studied here, the intermittent DRV step provides a deterministic
sub-grid correction whose residual errors can be traced to identifiable numerical ingredients rather than to irreducible
statistical modeling uncertainty.

\subsection{Reinjection of sharpened sub-cell information}

At each correction time the algorithm replaces a smeared numerical profile by a sharpened local profile reconstructed at
the current time. In that sense it does feed sub-cell information back into the resolved computation. We avoid calling
this ``exact backscatter,'' however, because the feedback still depends on finite-resolution DRV detection, plateau
sampling, and the approximate Newton closure. What the numerical results show is more concrete and, for the present
purposes, more important: on the severe-expansion, LeBlanc, and general-pattern benchmarks, this intermittent
reinjection can materially change the evolution of the resolved plateau states.

\section{Principal benchmarks}

We study two extended-time benchmarks that carry the main conceptual weight, together with two single-interface
general-pattern tests. Appendix~\ref{sec:hypdeg-appendix} records a separate illustrative near-vacuum problem that demonstrates the geometric
detection mechanism, and Appendix~\ref{sec:doublesod-appendix} records a noninteracting two-interface Double-Sod calculation.

\medskip
\noindent\textsf{Long-time severe expansion.}
We take
\[
\gamma=1.4,\qquad x_\ast=-0.2,\qquad (\rho_L,u_L,p_L)=(1,0,1),\qquad (\rho_R,u_R,p_R)=(10^{-4},0,10^{-4}),
\]
on the enlarged domain $[-1,2]$, with $N=900$ cells and final time $t=0.4$. The intermittent correction is applied every
$K=50$ steps. This test is designed to probe whether intermittent DRV correction maintains the star plateau over a
longer evolution than in a standard short-time benchmark.

\medskip
\noindent\textsf{Long-time LeBlanc.}
We take
\[
\gamma=\tfrac53,\qquad x_\ast=-\tfrac13,\qquad (\rho_L,u_L,p_L) =\bigl(1,0,\tfrac23\cdot 10^{-1}\bigr),\qquad
(\rho_R,u_R,p_R) =\bigl(10^{-3},0,\tfrac23\cdot 10^{-10}\bigr),
\]
on the enlarged domain $[-1.1,0.8]$, with $N=800$ cells and final time $t=1$. The intermittent correction is applied
every $K=3$ steps. This is the benchmark on which one-shot final-time DRV reconstruction fails and the intermittent step
becomes essential.

\medskip
\noindent\textsf{Long-time two-shock collision
($\mathbf{1}$-S/$\mathbf{2}$-C/$\mathbf{3}$-S).} We take the Toro Test~4 initial data~\cite{Toro2009}
\[
\begin{aligned}
\gamma&=1.4,\qquad x_\ast=0,\\
(\rho_L,u_L,p_L)&=(5.999,\;19.598,\;460.9),\\
(\rho_R,u_R,p_R)&=(5.992,\;{-6.196},\;46.10)
\end{aligned}
\]
on the enlarged domain $[-1,1]$, with $N=800$ cells and final time $t=0.07$. The intermittent correction is applied
every $K=10$ steps. Both the $1$-wave and $3$-wave are shocks; the DRV surrogates detect both as negative $\wdot$ spikes
(one on either side of the contact) and the one-step Newton closure automatically converges to the two-shock star-state
$\pstar\approx 1691.6$, $\ustar\approx 8.690$.

\medskip
\noindent\textsf{Long-time two-rarefaction
($\mathbf{1}$-R/$\mathbf{2}$-C/$\mathbf{3}$-R).} We take the Toro Test~2 initial data~\cite{Toro2009}
\[
\gamma=1.4,\qquad x_\ast=0,\qquad (\rho_L,u_L,p_L)=(1,\;{-2},\;0.4),\qquad (\rho_R,u_R,p_R)=(1,\;2,\;0.4),
\]
on the enlarged domain $[-1,1]$, with $N=800$ cells and final time $t=0.3$. The intermittent correction is applied every
$K=20$ steps. Both the $1$-wave and $3$-wave are rarefactions, creating a symmetric near-vacuum expansion with
$\pstar\approx 1.89\times 10^{-3}$. There are no shocks; the DRV surrogates detect both rarefaction fans via the $\zdot$
support and $u_x$ positivity.

Table~\ref{tab:setup} summarizes the resolutions and correction frequencies used in the principal single-interface
computations. The separate Double-Sod appendix uses $[-1,1]$, $N=1200$, $t=0.1$, and $K=20$.

\begin{table}[ht]
\centering
\renewcommand{\arraystretch}{1.15}
\caption{{\footnotesize Benchmark setup for the principal single-interface
  computations. The long-time severe-expansion run uses $K=50$ (36 corrections over 1812 steps); the long-time LeBlanc
  run uses $K=3$ (976 corrections over 2931 steps). The two-shock and two-rarefaction cases test generalization beyond
  the standard $1$-R/$2$-C/$3$-S pattern.}}
\label{tab:setup}
\begin{tabular}{llccccc}
\toprule
\textsf{Benchmark} & \textsf{Pattern} & \textsf{Domain} & \textsf{$t_{\mathrm{fin}}$}
  & \textsf{$N$} & \textsf{$K$} & \textsf{Steps} \\
\midrule
Long-time severe expansion & 1-R/2-C/3-S & $[-1,2]$ & $0.4$ & $900$ & $50$
  & $1812$ \\
Long-time LeBlanc & 1-R/2-C/3-S & $[-1.1,0.8]$ & $1.0$ & $800$ & $3$
  & $2931$ \\
Two-shock collision & 1-S/2-C/3-S & $[-1,1]$ & $0.07$ & $800$ & $10$
  & $3357$ \\
Two-rarefaction & 1-R/2-C/3-R & $[-1,1]$ & $0.3$ & $800$ & $20$
  & $1320$ \\
\bottomrule
\end{tabular}
\end{table}

\section{Long-time severe expansion}
\label{sec:severe}

The long-time severe-expansion problem is the benchmark on which the LES-like interpretation
(Section~\ref{sec:les-analogy}) is clearest. A one-shot final-time DRV reconstruction of the uncorrected WENO-5/HLLC run
already places the waves reasonably well, but the raw star plateau remains contaminated by wall-heating drift.
Intermittent DRV correction every $K=50$ steps removes that contamination almost completely.

The numerical summary is given in Table~\ref{tab:main-results}. The median left-star plateau errors in velocity and
pressure drop from $5.74\times 10^{-2}$ and $2.39\times 10^{-4}$ for the one-shot final-time reconstruction to
$2.41\times 10^{-13}$ and $1.75\times 10^{-15}$ under intermittent correction. The final contact and shock location
errors drop from $6.89\times 10^{-4}$ and $1.86\times 10^{-3}$ to the $10^{-13}$ level.

This is the deterministic sub-grid correction mechanism at work: each intermittent correction samples the uncorrupted
far-field plateaus, applies one Newton step of the distributed iteration (Section~\ref{sec:distributed-newton}), and
injects the resulting sharp state back into the resolved solution. The far-field states are always clean because they
sit in constant regions untouched by any wave. The star-state, which the uncorrected WENO-5 scheme progressively
contaminates through numerical diffusion, is repeatedly reset to its exact value.

\begin{figure}[ht]
\centering
\includegraphics[width=\textwidth]%
  {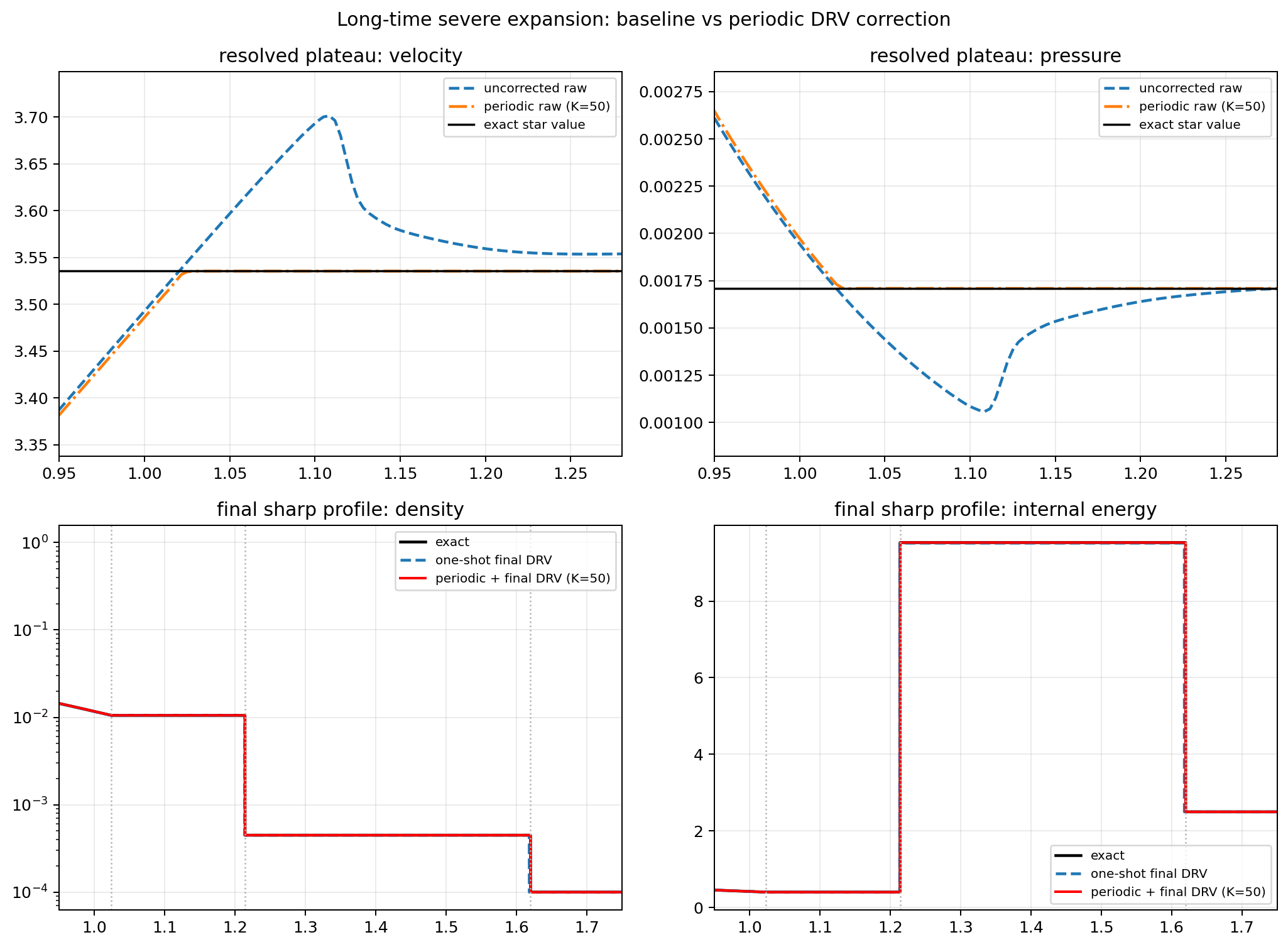}
\caption{{\footnotesize Long-time severe expansion on $N=900$ cells at $t=0.4$.
  Top row: resolved plateau velocity and pressure for the uncorrected run and the intermittently corrected run ($K=50$),
  together with the exact star-state values. Bottom row: final sharp density and internal-energy profiles. Intermittent
  DRV correction removes the plateau drift visible in the uncorrected run, producing a final sharp reconstruction that
  is effectively exact.}}
\label{fig:severe}
\end{figure}

Figure~\ref{fig:severe} makes the mechanism visible. The top row shows the resolved plateau variables: the uncorrected
run drifts measurably away from the exact star-state values, while the intermittent run sits on top of them to plotting
accuracy. The bottom row shows the final sharp reconstruction: the one-shot final-time reconstruction is already
respectable, but the intermittent run is effectively exact.

\section{Long-time LeBlanc}

The long-time LeBlanc problem is the stronger benchmark. Here one-shot final-time DRV reconstruction is no longer
sufficient. The uncorrected run reaches $t=1$, but the final sharp reconstruction built from that run places the shock
incorrectly by
\[
|X_s-X_s^{\mathrm{ex}}| = 2.71\times 10^{-1},
\]
and its sharp-profile $L^1$ velocity error is $1.90\times 10^{-1}$.

\begin{figure}[ht]
\centering
\includegraphics[width=\textwidth]%
  {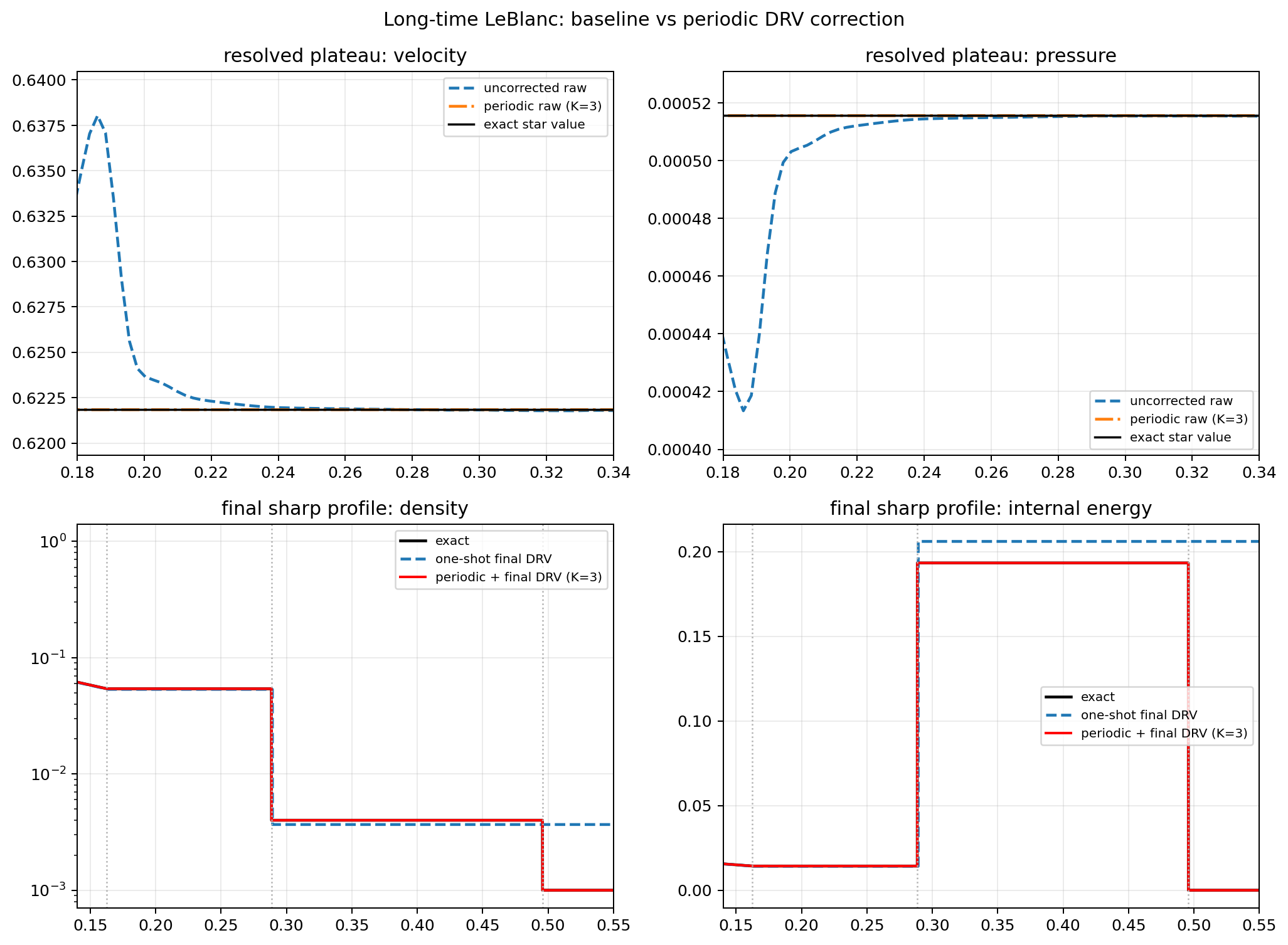}
\caption{{\footnotesize Long-time LeBlanc on $N=800$ cells at $t=1$.  Top row:
  resolved plateau velocity and pressure for the uncorrected run and the intermittently corrected run ($K=3$), together
  with the exact star-state values. Bottom row: final sharp density and internal-energy profiles. This is the benchmark
  on which one-shot final-time DRV reconstruction fails and intermittent correction becomes essential.}}
\label{fig:leblanc}
\end{figure}

Intermittent correction changes that conclusion completely. With correction frequency $K=3$, the contact and shock are
recovered to plotting accuracy, the median star-plateau errors drop to machine precision, and the sharp-profile $L^1$
errors fall to the $10^{-5}$--$10^{-7}$ range (Table~\ref{tab:main-results}). Figure~\ref{fig:leblanc} shows the effect:
the uncorrected final-time-only reconstruction misses the right shock and carries a visibly wrong internal-energy
plateau, whereas the intermittently corrected run reproduces the exact geometry almost perfectly.

A noteworthy feature is that the correction frequency matters sharply. Table~\ref{tab:leblanc-sweep} shows the observed
transition. Sparse intermittent correction ($K\ge 4$) does not help this benchmark; in fact, it causes the time step to
collapse before $t=1$ is reached. By contrast, $K=3$ reaches $t=1$ and restores exact wave geometry. The non-monotonic
dependence of $t_{\mathrm{end}}$ on $K$ in Table~\ref{tab:leblanc-sweep} (e.g.\ $K=50$ reaches $t\approx 0.16$ while
$K=10$ collapses at $t\approx 0.08$) reflects the interaction between the correction frequency and the accumulation of
wall-heating error: less frequent but still insufficiently frequent correction can temporarily stabilize the plateau
long enough for the solver to advance further before the eventual collapse.

Intermittent correction is therefore not automatically stabilizing. On this benchmark it must be applied often enough to
prevent the plateau drift from leaving the recoverable regime.

\section{Two-shock collision ($1$-S/$2$-C/$3$-S)}

The two-shock collision benchmark demonstrates that the intermittent DRV mechanism generalizes beyond the standard
$1$-rarefaction/$2$-contact/$3$-shock pattern. Here both the $1$-wave and $3$-wave are shocks; no rarefaction is
present. The DRV surrogates detect both shocks as negative $\wdot$ spikes---one to the left and one to the right of the
$\sdot$ contact spike---and the one-step Newton closure automatically converges to the two-shock star-state.

\begin{figure}[ht]
\centering
\includegraphics[width=\textwidth]%
  {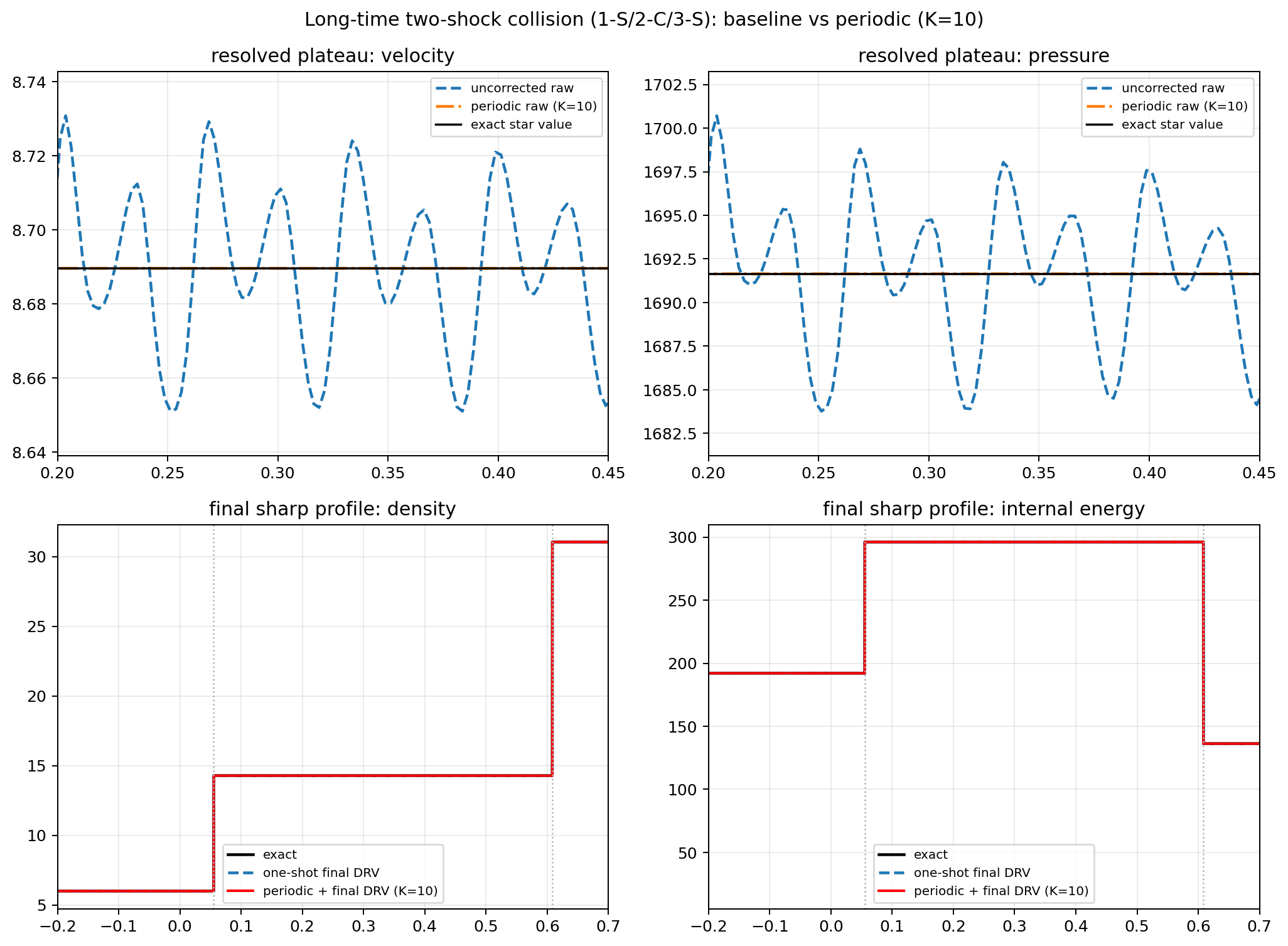}
\caption{{\footnotesize Long-time two-shock collision ($1$-S/$2$-C/$3$-S) on
  $N=800$ cells at $t=0.07$. Top row: resolved plateau velocity and pressure. The uncorrected run (blue dashed) exhibits
  wall-heating oscillations; intermittent correction $K=10$ (orange dash-dot) maintains the exact star values. Bottom
  row: final sharp density and internal-energy profiles, both effectively exact.}}
\label{fig:twoshock}
\end{figure}

The plateau stabilization is dramatic. Figure~\ref{fig:twoshock} makes the effect visible. The uncorrected WENO-5
solution develops wall-heating oscillations in the star-state velocity and pressure with amplitudes of $\Delta u \approx
1.6 \times 10^{-2}$ and $\Delta p \approx 3.0$. Intermittent correction with $K=10$ drives these errors to $5.2 \times
10^{-8}$ and $9.4 \times 10^{-5}$, respectively---an improvement of five orders of magnitude in velocity and four in
pressure (Table~\ref{tab:main-results}). The final sharp reconstruction is excellent from both the corrected and
uncorrected runs, confirming that the principal effect of intermittent correction here is plateau stabilization.

\section{Two-rarefaction expansion ($1$-R/$2$-C/$3$-R)}

\begin{figure}[ht]
\centering
\includegraphics[width=\textwidth]%
  {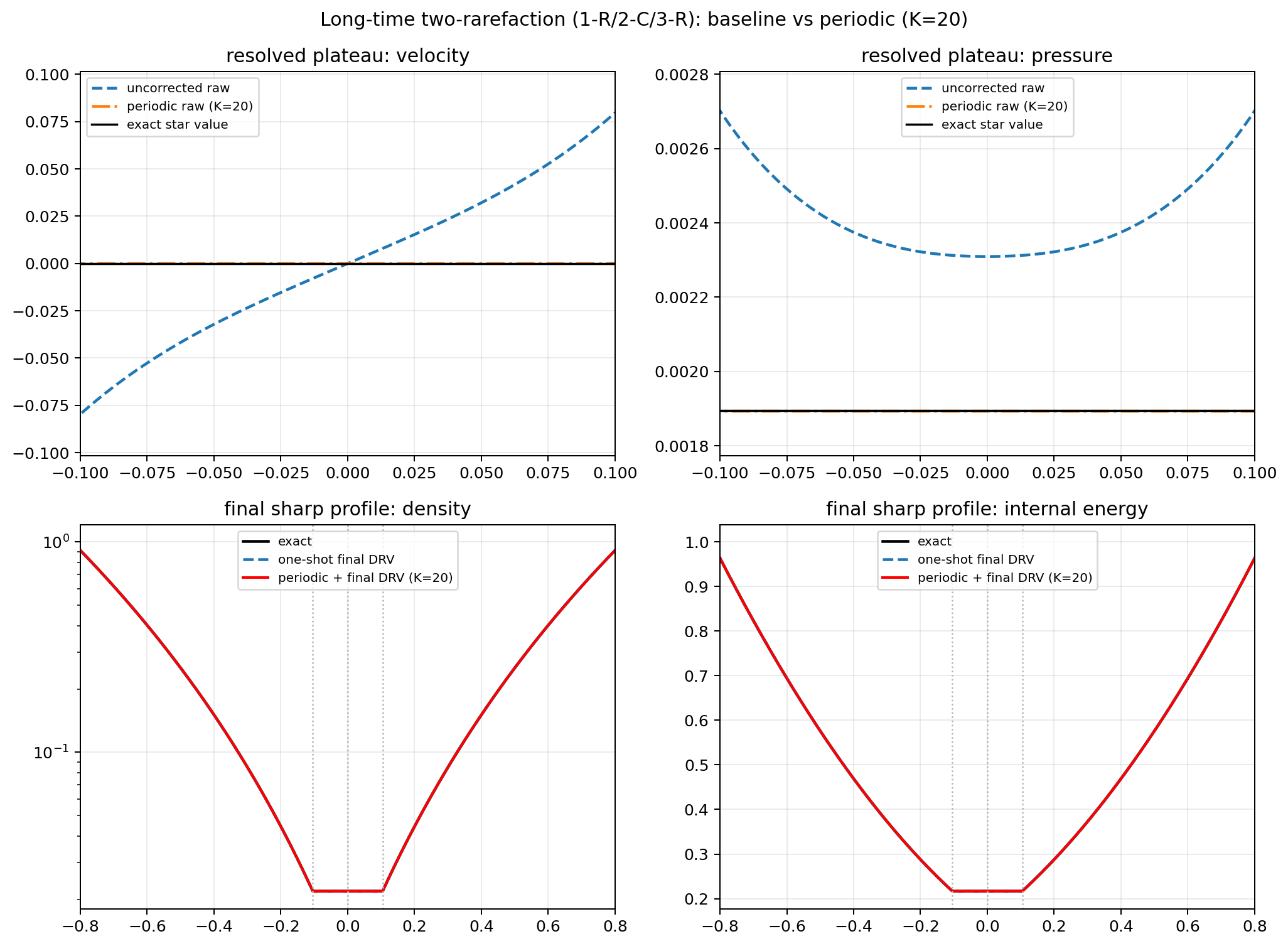}
\caption{{\footnotesize Long-time two-rarefaction expansion ($1$-R/$2$-C/$3$-R)
  on $N=800$ cells at $t=0.3$. Top row: the uncorrected velocity plateau (blue dashed) drifts linearly from $-0.075$ to
  $+0.075$ around the exact $\ustar=0$; intermittent correction $K=20$ (orange dash-dot) holds $u=0$ to machine
  precision. The pressure plateau is similarly corrected. Bottom row: final sharp profiles, both overlaying the exact
  solution.}}
\label{fig:tworar}
\end{figure}

The two-rarefaction benchmark is the complementary generalization. Figure~\ref{fig:tworar} shows the resulting plateau
stabilization and final sharp profiles. Both the $1$-wave and $3$-wave are rarefactions, no shocks are present, and the
star region is a symmetric near-vacuum with $\pstar \approx 1.89\times 10^{-3}$. The DRV surrogates detect both
rarefaction fans via the $\zdot$ support and the positivity of $\partial_x u$.

Intermittent correction with $K=20$ drives the star-plateau velocity error from $3.4\times 10^{-2}$ to machine precision
($5.5\times 10^{-18}$), and the pressure error from $4.9\times 10^{-4}$ to $1.2\times 10^{-6}$
(Table~\ref{tab:main-results}). This is a sixteen-order improvement in velocity. As in the two-shock case, the final
sharp reconstruction is already excellent without intermittent correction, so the headline improvement is again in the
evolving plateau.

\section{Combined numerical summary}

The principal quantitative comparison is collected in Table~\ref{tab:main-results}. Here $\mathrm{med}\,|u-\ustar|$ and
$\mathrm{med}\,|p-\pstar|$ are computed over the cell centers in the exact left-star interval
$$(X_{\mathrm{rt}}+2\,\dx,\,X_c-2\,\dx),$$
so they are intentionally local plateau metrics. A small median plateau error therefore does not by itself control the
far-right wave placement, which is why the long-time LeBlanc final-time-only row can show a small local plateau error
together with a catastrophic right-wave error. For the two $1$-R/$2$-C/$3$-S benchmarks (severe expansion and
LeBlanc), the headline improvements are in the evolving plateau and wave positions; for the LeBlanc problem, the
improvement is qualitative: one-shot reconstruction fails while intermittent correction succeeds. For the two
general-pattern benchmarks (two-shock and two-rarefaction), the intermittent correction confirms that the DRV
mechanism is pattern-agnostic, with plateau improvements of four to sixteen orders of magnitude.

\begin{table}[ht]
\centering
\footnotesize
\renewcommand{\arraystretch}{1.15}
\caption{{\footnotesize Principal numerical comparison.  ``Final-time only''
  means a single DRV reconstruction applied at the end of an otherwise uncorrected WENO-5/HLLC run. ``Intermittent''
  means every-$K$-step DRV correction inserted during the evolution, using the approximate one-step local closure and no
  exact local Riemann solve. Here $X_{3,\mathrm{in}}$ denotes the inner edge of the right acoustic family: the shock
  position when the $3$-wave is a shock and the rarefaction tail when the $3$-wave is a rarefaction.}}
\label{tab:main-results}
\begin{tabular}{>{\raggedright\arraybackslash}p{2.2cm}
  >{\raggedright\arraybackslash}p{1.7cm} c c c c}
\toprule
\textsf{Benchmark} & \textsf{Method} & $\mathrm{med}\,|u-\ustar|$ & $\mathrm{med}\,|p-\pstar|$ &
$|X_c-X_c^{\mathrm{ex}}|$
  & $|X_{3,\mathrm{in}}-X_{3,\mathrm{in}}^{\mathrm{ex}}|$ \\
\midrule
Severe expansion & final-time only & $5.74\times 10^{-2}$ & $2.39\times 10^{-4}$ & $6.89\times 10^{-4}$
  & $1.86\times 10^{-3}$ \\

Severe expansion & intermittent $K=50$ & $2.41\times 10^{-13}$ & $1.75\times 10^{-15}$ & $1.49\times 10^{-13}$ &
$1.68\times 10^{-13}$ \\[1mm]

Long-time LeBlanc & final-time only & $3.23\times 10^{-4}$ & $2.52\times 10^{-6}$ & $7.33\times 10^{-4}$
  & $2.71\times 10^{-1}$ \\

Long-time LeBlanc & intermittent $K=3$ & $2.47\times 10^{-13}$ & $1.08\times 10^{-15}$ & $0$ & $0$ \\[1mm]

Two-shock & final-time only & $1.57\times 10^{-2}$ & $2.97$ & $4.97\times 10^{-9}$
  & $4.85\times 10^{-8}$ \\

Two-shock & intermittent $K=10$ & $5.23\times 10^{-8}$ & $9.40\times 10^{-5}$ & $4.99\times 10^{-9}$ & $4.87\times
10^{-8}$ \\[1mm]

Two-rarefaction & final-time only & $3.40\times 10^{-2}$ & $4.89\times 10^{-4}$ & $4.66\times 10^{-16}$
  & $1.25\times 10^{-6}$ \\

Two-rarefaction & intermittent $K=20$ & $5.45\times 10^{-18}$ & $1.18\times 10^{-6}$ & $0$
  & $1.53\times 10^{-6}$ \\
\bottomrule
\end{tabular}
\end{table}

\begin{table}[ht]
\centering
\renewcommand{\arraystretch}{1.15}
\caption{{\footnotesize Correction-frequency sweep for the long-time LeBlanc
  benchmark ($N=800$, $t=1$). For $K\ge 4$ the time step collapses before final time. The threshold $K=3$ already
  restores the exact contact and shock positions to plotting accuracy. The non-monotonic dependence of
  $t_{\mathrm{end}}$ on $K$ is discussed in the text.}}
\label{tab:leblanc-sweep}
\begin{tabular}{cccccc}
\toprule
\textsf{$K$} & \textsf{Completed?} & \textsf{$t_{\mathrm{end}}$} & \textsf{Corrections} & $|X_c-X_c^{\mathrm{ex}}|$
  & $|X_{3,\mathrm{in}}-X_{3,\mathrm{in}}^{\mathrm{ex}}|$ \\
\midrule
$0$ & yes & $1.0000$ & $0$
  & $7.33\times 10^{-4}$ & $2.71\times 10^{-1}$ \\
$50$ & no  & $0.1612$ & $5$   & --- & --- \\
$20$ & no  & $0.1899$ & $13$  & --- & --- \\
$10$ & no  & $0.0806$ & $11$  & --- & --- \\
$5$  & no  & $0.0854$ & $23$  & --- & --- \\
$4$  & no  & $0.0755$ & $25$  & --- & --- \\
$3$  & yes & $1.0000$ & $976$ & $0$ & $0$ \\
\bottomrule
\end{tabular}
\end{table}

\subsection{Wall-clock cost}

Because part of the appeal of the method is its cost-to-effect ratio, it is useful to record representative timing
information. Table~\ref{tab:timings} reports single-run wall-clock timings on the present machine for the current
pure-Python prototype. The absolute times are machine dependent, so the main point is the relative overhead of inserting
the intermittent DRV correction.

The computational message is favorable but not uniform. When the correction is infrequent, the added wall-clock cost is
small: about $10\%$ on the two-shock benchmark and about $5\%$ on the two-rarefaction benchmark. On the long-time
severe-expansion run, the intermittently corrected computation is even slightly faster because the correction reduces
the total number of Euler steps. The demanding long-time LeBlanc benchmark is the expensive case: the successful regime
requires aggressive correction every $K=3$ steps, and the wall-clock time rises by about $84\%$. Even there, however,
the runtime penalty remains below a factor of two in this unoptimized Python prototype, while the qualitative gain in
recoverability is decisive.

\begin{table}[ht]
\centering
\renewcommand{\arraystretch}{1.15}
\caption{{\footnotesize Representative wall-clock timings for the current
  pure-Python prototype on the present machine. ``Baseline'' means the uncorrected WENO-5/HLLC run plus the final sharp
  reconstruction; ``intermittent'' means the same code with DRV correction inserted every $K$ steps.}}
\label{tab:timings}
\begin{tabular}{lcccc}
\toprule
\textsf{Benchmark} & \textsf{$K$} & \textsf{Baseline (s)} & \textsf{Intermittent (s)}
  & \textsf{Factor} \\
\midrule
Long-time severe expansion
  & $50$ & $8.88$ & $8.26$ & $0.93\times$ \\
Long-time LeBlanc
  & $3$ & $10.76$ & $19.79$ & $1.84\times$ \\
Two-shock collision
  & $10$ & $11.28$ & $12.42$ & $1.10\times$ \\
Two-rarefaction
  & $20$ & $4.40$ & $4.60$ & $1.05\times$ \\
\bottomrule
\end{tabular}
\end{table}

\section{Discussion}

\subsection{Deterministic versus statistical sub-grid structure}

The success of the DRV correction highlights a fundamental asymmetry between compressible wave dynamics and turbulence.
In turbulence, the sub-grid scales are chaotic: they exhibit sensitive dependence on initial conditions, the energy
cascade is statistical, and no deterministic closure is available.

In the one-dimensional local Riemann packets considered here, the sub-cell structure is much more rigid. Between
adjacent waves the state is constant, and the intermediate states are determined by the local Riemann structure once the
outer states and wave types are known. The DRV surrogates provide the wave positions, while the Newton update supplies a
cheap Riemann-informed closure for the star state. The residual error then comes from numerical detection, plateau
sampling, and finite correction frequency, rather than from a turbulence-style modeling ansatz.

\subsection{Beyond the standard pattern}

The two-shock collision and two-rarefaction benchmarks demonstrate that the DRV detection and Newton closure are
genuinely pattern-flexible. No case-specific logic is needed: the DRV surrogates automatically identify shock spikes
(via $\wdot$) and rarefaction fans (via $\zdot$ support and $u_x$ positivity), and the one-step Newton closure returns
the correct wave types as a by-product. The principal single-interface computations in this paper therefore cover three
distinct classical patterns: $1$-R/$2$-C/$3$-S, $1$-S/$2$-C/$3$-S, and $1$-R/$2$-C/$3$-R. Appendix~\ref{sec:doublesod-appendix} adds a
noninteracting two-interface Double-Sod calculation whose right packet realizes the remaining local $1$-S/$2$-C/$3$-R
configuration.

\subsection{Scope and limitations}

The principal body of the paper treats single local Riemann packets with known interface positions, and Appendix~\ref{sec:doublesod-appendix} adds
a finite collection of two noninteracting local packets with fixed windows. The present paper therefore does \emph{not}
claim a general-purpose LES-style model for arbitrary Euler data, nor does it yet address post-interaction wave dynamics
or genuinely smooth non-Riemann initial data.

What it does show is that, within the present one-dimensional Riemann-packet setting, the intermittent DRV mechanism can
function as a genuine evolution-time correction: it can maintain or recover hidden wave geometry strongly enough to
change the outcome of the computation. The extension to interacting wave packets and to multidimensional Euler data
remains the natural next step.

\section{Conclusion}

On the benchmarks studied here, intermittent DRV correction changes the computation itself, not merely the appearance of
its final plot.

On the long-time severe-expansion problem it drives the evolving intermediate states to machine precision and makes the
final contact and shock locations essentially exact. On the long-time LeBlanc problem the statement is stronger: without
intermittent correction, final-time DRV reconstruction no longer succeeds, whereas correction every three steps restores
an almost exact solution. On the two-shock collision and two-rarefaction benchmarks, the same detector-and-Newton
mechanism extends beyond the standard $1$-R/$2$-C/$3$-S pattern, while Appendix~\ref{sec:doublesod-appendix} shows that the method is not tied to a
single global similarity center.

The LES comparison developed in Section~\ref{sec:les-analogy} is best read as an interpretive analogy. In practical
terms, the method adds filtered DRV differences, local state sampling, a short Newton correction, and conservative
remapping to an otherwise standard fixed-grid solver. On the problems studied here, that small addition has a
disproportionate effect: from grids that would ordinarily seem far too coarse, it recovers sharp solutions that are
nearly exact. Table~\ref{tab:timings} shows that this gain comes with only a small-to-moderate wall-clock penalty in the
present Python prototype, which strengthens the case for trying the method in an existing 1D code.

\appendix

\section{A noninteracting two-interface Double-Sod illustration}
\label{sec:doublesod-appendix}
\begin{figure}[H]
\centering
\includegraphics[width=0.8\textwidth]%
  {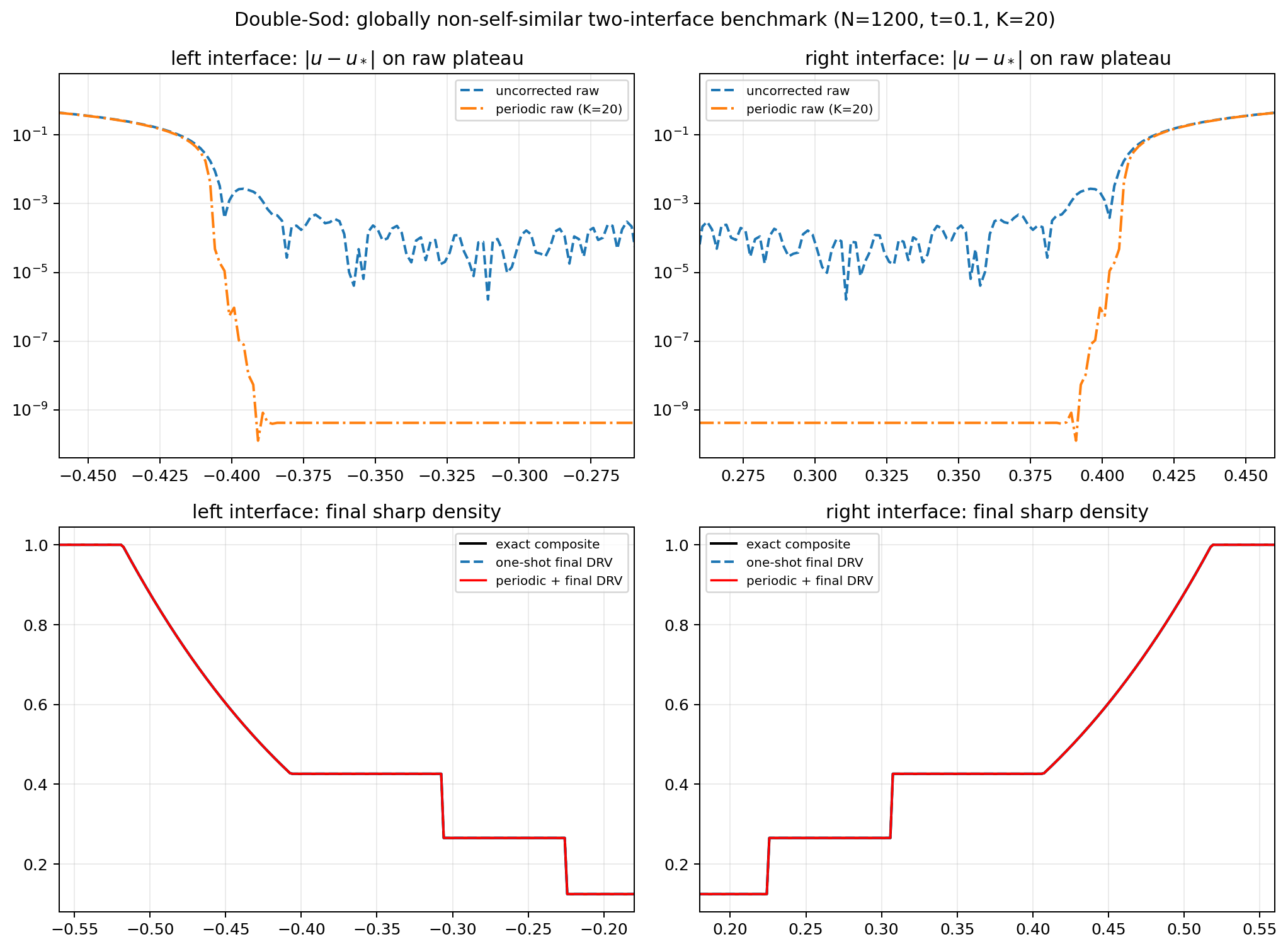}
\caption{{\footnotesize Double-Sod on $N=1200$ cells at $t=0.1$.  The initial
  data contain two interfaces at $x=\pm 0.4$, so the global solution is not self-similar. Top row: pointwise
  $|u-\ustar|$ errors on the raw plateau for the uncorrected run and the intermittently corrected run ($K=20$), shown
  separately for the left and right star regions. Bottom row: final sharp density around the left and right interfaces.
  The intermittent correction can be applied independently on the two half-domains because the two local wave packets
  remain disjoint up to $t=0.1$.}}
\label{fig:doublesod}
\end{figure}

To show that the intermittent DRV mechanism is not tied to a single global similarity center, we record a short
two-interface ``Double-Sod'' computation.
The initial data are
\[
(\rho,u,p)=
\begin{cases}
(1,0,1), & x<-0.4,\\
(0.125,0,0.1), & -0.4<x<0.4,\\
(1,0,1), & x>0.4,
\end{cases}
\]
on $[-1,1]$, with $N=1200$ cells and final time $t=0.1$. Intermittent correction is applied every $K=20$ steps,
separately on the two local windows $[-1,0]$ and $[0,1]$. Up to $t=0.1$ the two local wave packets remain disjoint, so
the exact solution is the superposition of two nonoverlapping local Riemann solutions. The left packet has the standard
$1$-R/$2$-C/$3$-S structure, while the right packet realizes the remaining local $1$-S/$2$-C/$3$-R configuration.

This appendix is illustrative rather than definitive: it does not yet treat post-interaction dynamics.
Figure~\ref{fig:doublesod} summarizes the computation, but it also shows something useful: the intermittent DRV
mechanism can be run on more than one local packet at a time. Taking the maximum over the two local interfaces,
intermittent correction reduces the raw plateau velocity error from $1.62\times 10^{-4}$ to $4.15\times 10^{-10}$ and
the raw plateau pressure error from $5.98\times 10^{-5}$ to $1.09\times 10^{-9}$. The composite sharp-profile $L^1$
density error also drops from $3.35\times 10^{-7}$ to $1.13\times 10^{-8}$.

\section{Illustrative sub-cell anatomy: the
  hyperbolic-degeneracy problem}
\label{sec:hypdeg-appendix}
\begin{figure}[ht]
\centering
\includegraphics[width=\textwidth]%
  {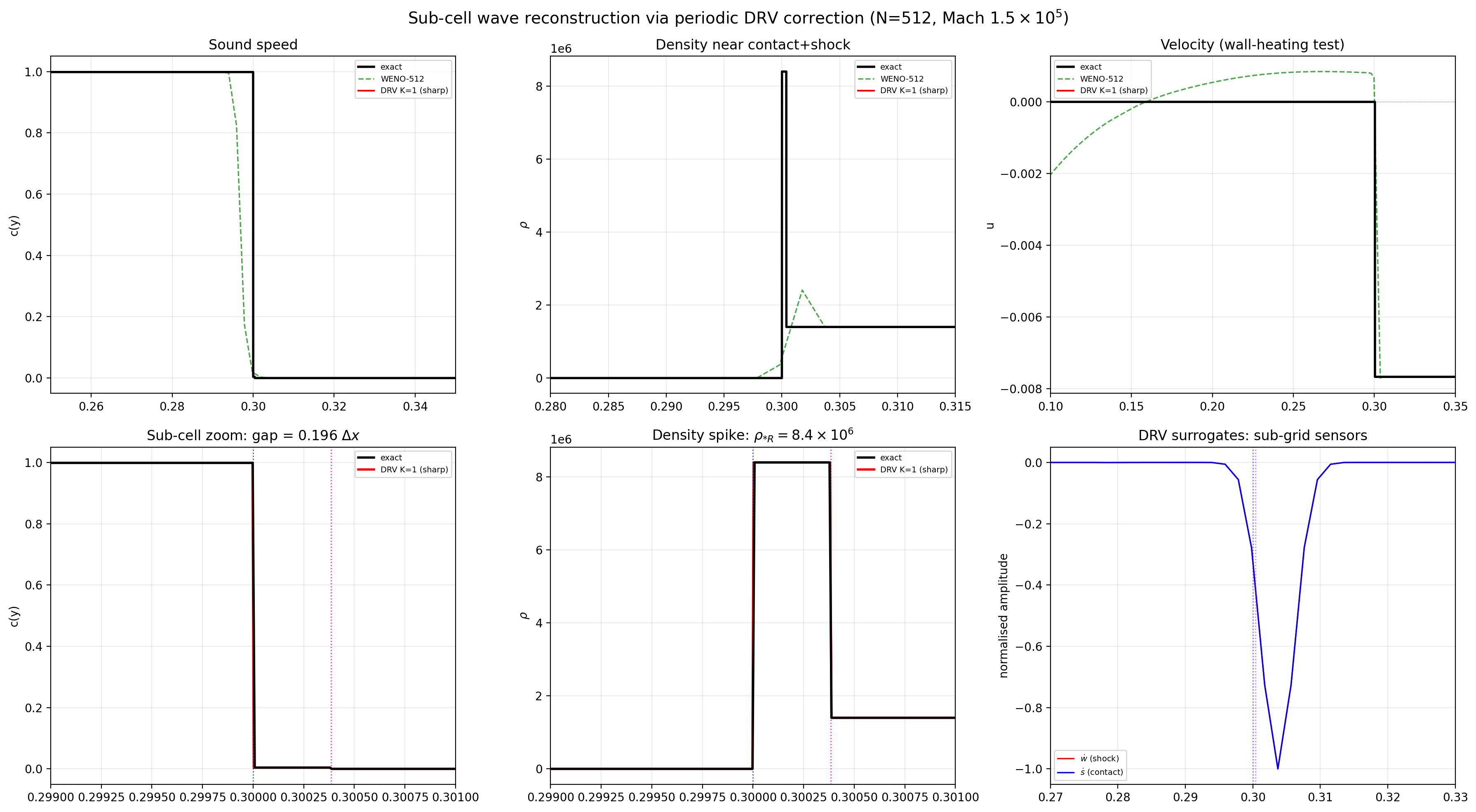}
\caption{{\footnotesize Illustrative benchmark: the hyperbolic-degeneracy
  near-vacuum problem on $N=512$ cells (Mach $1.53\times 10^5$, contact-shock gap $=0.196\,\dx$). The DRV surrogates
  detect the contact ($\sdot$, blue) and shock ($\wdot$, red) as distinct features in different characteristic families,
  and the sharp reconstruction recovers the exact sub-cell anatomy.}}
\label{fig:illustrative-hypdeg}
\end{figure}

This appendix includes Figure~\ref{fig:illustrative-hypdeg} to illustrate the geometric mechanism that underlies the DRV
detection. The benchmark uses $\gamma=1.4$, $x_\ast=0.3$, and Riemann data producing a Mach $1.53\times 10^5$ shock with
a contact-shock separation of $0.196\,\dx$ on $N=512$ cells. Even when the contact and shock lie inside a sub-cell gap
of $0.196\,\dx$, the DRV surrogates---which diagonalize the characteristic information that the numerical scheme has
blended---distinguish their locations and the final sharp reconstruction recovers the hidden anatomy.

On this particular benchmark a one-shot final-time DRV reconstruction already solves the case very well, so the
intermittent correction does not play a decisive role. The benchmark is therefore included only to illustrate the
detection mechanism, not to argue for the necessity of intermittent correction.

\section*{Acknowledgments}
This work was supported by NSF grant DMS-2307680.


\begin{thebibliography}{99}

\bibitem{DAquilaHelenbrookMazaheri2020}
L.~D'Aquila, B.~T.~Helenbrook, and A.~Mazaheri.
\newblock High-order shock fitting with finite element methods.
\newblock In \emph{AIAA Aviation 2020 Forum}, AIAA Paper
  2020-3047, 2020.

\bibitem{JiangShu1996}
G.-S.~Jiang and C.-W.~Shu.
\newblock Efficient implementation of weighted {ENO} schemes.
\newblock \emph{J.~Comput.\ Phys.}, 126(1):202--228, 1996.

\bibitem{LeVequeShyue1995}
R.~J.~LeVeque and K.-M.~Shyue.
\newblock One-dimensional front tracking based on high resolution
  wave propagation methods.
\newblock \emph{SIAM J.~Sci.\ Comput.}, 16(2):348--377, 1995.

\bibitem{Pope2000}
S.~B.~Pope.
\newblock \emph{Turbulent Flows}.
\newblock Cambridge University Press, Cambridge, 2000.

\bibitem{ReisnerSerencsaShkoller2013}
J.~Reisner, J.~Serencsa, and S.~Shkoller.
\newblock A space-time smooth artificial viscosity method for
  nonlinear conservation laws.
\newblock \emph{J.~Comput.\ Phys.}, 235:912--933, 2013.

\bibitem{ShkollerDRV2026}
S.~Shkoller.
\newblock Sub-cell wave reconstruction from differentiated
  {R}iemann variables.
\newblock Preprint,
  \href{https://arxiv.org/abs/2603.16830}{arXiv:2603.16830}, 2026.

\bibitem{ShuOsher1989}
C.-W.~Shu and S.~Osher.
\newblock Efficient implementation of essentially non-oscillatory
  shock-capturing schemes, {II}.
\newblock \emph{J.~Comput.\ Phys.}, 83(1):32--78, 1989.

\bibitem{Toro2009}
E.~F.~Toro.
\newblock \emph{Riemann Solvers and Numerical Methods for Fluid
  Dynamics: A Practical Introduction}.
\newblock Springer, Berlin, third edition, 2009.

\bibitem{ToroSpruceSpeares1994}
E.~F.~Toro, M.~Spruce, and W.~Speares.
\newblock Restoration of the contact surface in the
  {HLL}-{R}iemann solver.
\newblock \emph{Shock Waves}, 4(1):25--34, 1994.

\bibitem{WitteveenKorenBakker2007}
J.~A.~S.~Witteveen, B.~Koren, and P.~G.~Bakker.
\newblock An improved front tracking method for the Euler
  equations.
\newblock \emph{J.~Comput.\ Phys.}, 224(2):712--728, 2007.

\end{thebibliography}
\end{document}